\documentclass[aps,pre,twocolumn,twoside, amsmath,amssymb, superscriptaddress,
 aps,bbm,floatfix,showpacs]{revtex4-1}

\usepackage{xcolor}
\usepackage{graphicx}
\usepackage{subfigure}
\usepackage{dcolumn}
\usepackage{bm}
\usepackage{framed}
\usepackage{hyperref}
\usepackage{dsfont}
\usepackage[normalem]{ulem}
\usepackage{nicefrac}

\begin{document}

\title{Heat leakage in overdamped harmonic systems}
\author{Dominic Arold}
\affiliation{Department of Physics, Friedrich-Alexander-Universit\"at Erlangen-N\"urnberg, D-91058 Erlangen, Germany}
 \author{Andreas Dechant}
 \affiliation{Department of Physics, Graduate School of Science, Kyoto University, Kyoto 606-8502, Japan}
\author{Eric Lutz}
\affiliation{Department of Physics, Friedrich-Alexander-Universit\"at Erlangen-N\"urnberg, D-91058 Erlangen, Germany}

\begin{abstract}
We investigate the occurrence of heat leakages in overdamped Brownian harmonic systems. We exactly compute the underdamped and overdamped stochastic heats exchanged with the  bath for a sudden frequency or temperature switch. We show that the underdamped heat reduces to the corresponding overdamped expression in the limit of large friction for the isothermal process. However, we establish that this is not the case for the isochoric transformation. We  microscopically derive the additionally generated heat leakage and  relate its origin to the initial relaxation of the velocity of the system. Our results highlight the limitations of the overdamped approximation for the evaluation of the stochastic heat in systems with changing bath temperature.
\end{abstract}

\maketitle

\section{Introduction}
Stochastic thermodynamics offers a general framework for the study of  the thermodynamic properties of small systems whose dynamics is dominated by thermal fluctuations \cite{sek10,sei12}. It  successfully extends the concepts of macroscopic thermodynamics, such as work, heat, energy  and entropy, to the level of single random trajectories \cite{sek98,sei05}. This framework  has been widely  used  in theoretical and experimental investigations \cite{sek10,sei12,sek98,sei05,jar11,cil13}. An important application of stochastic thermodynamics is the analysis of Brownian heat engines that cyclically transform heat into mechanical work. Two broad classes of stochastic heat engines are usually distinguished: i) motors  with a static periodic potential and a spatially varying temperature \cite{but87,kam87,lan88}, and ii) motors with a time-dependent harmonic potential and a temporally modulated temperature \cite{sch07,esp09,esp10}. Examples of the second type have recently been realized experimentally using a colloidal particle trapped in a harmonic  optical potential \cite{bli12,mar15}. While most investigations have primarily focused  on the overdamped regime corresponding to large friction, studies in the underdamped limit of small friction have been performed for the two  different classes of stochastic engines \cite{bla98,dec15}.

The formalism of stochastic thermodynamics has been developed both for underdamped and overdamped dynamics \cite{sek10,sei12}. The main difference between the two is the definition of heat that includes a kinetic contribution in the underdamped regime, while it only depends on the confining potential in the overdamped limit. However, an asymptotic analysis of the Klein-Kramers equation  for the first class of Brownian heat engines has shown that the kinetic contribution does not vanish in the limit of large friction \cite{hon00}. The  heat flow  that results from the relaxation of the kinetic energy leads to a heat leakage that seriously limits the efficiency of the heat engine \cite{hon00,sek00,ben08}. A similar phenomenon has been observed for the second type of stochastic heat engines \cite{sch07,dec16a}, but has not been examined in detail to our knowledge.

Our aim in this paper is to investigate the appearance of heat leakages in harmonic systems. We specifically consider isothermal (constant temperature) and isochoric (constant frequency) processes that constitute two essential steps of a harmonic Brownian heat engine cycle. We exactly calculate   the underdamped and the overdamped heats for the two processes for the case of a sudden variation of frequency and temperature, respectively. We find that the underdamped heat reduces to the corresponding overdamped expression in the limit of large friction for the isothermal process, as expected. However, we show that this is not the case for the isochoric process and explicitly derive the non-vanishing heat leakage. 

\section{Heat for a harmonic  particle}
\label{Ch.HeatWorkforBP}
We consider a  Brownian particle with position $x$, velocity $v$ and mass $m$ confined in the harmonic  potential $V(x) = m\omega^2 x^2/2$ with frequency $\omega$. Its underdamped evolution is  described by the Langevin equation \cite{ris89},
\begin{equation}
\dot{v} = -\gamma v -\omega^2 x + \sqrt{\dfrac{2\gamma kT}{m}} F(t), \quad\dot{x} = v,
\label{LangevinParaPot}
\end{equation}
where $F(t)$ is a centered Gaussian white noise obeying $\langle F(t)F(t')\rangle = \delta(t-t')$, $k$ the Boltzmann constant, $T$ the temperature of the bath and $\gamma$ the friction coefficient. Because of the linearity of Eq.~\eqref{LangevinParaPot}, the dynamics  may be equivalently described in terms of  the position and velocity variances,  $\sigma_x = \langle x^2\rangle$ and $\sigma_v = \langle v^2\rangle$, leading to,
\begin{eqnarray}
\dot \sigma_v &=& -2\gamma \sigma_v -\omega^2 \dot \sigma_x + \dfrac{2\gamma k T}{m}, \\
\ddot \sigma_x &=& 2 \sigma_v - \gamma \dot \sigma_x - 2\omega^2 \sigma_x. 
\end{eqnarray}
In the overdamped limit, $\dot \sigma_v=0$,  the above equations decouple and  the Langevin equation  reduces to,
\begin{align}
\dot \sigma_x= -2\dfrac{\omega^2}{\gamma}\sigma_x + \frac{2kT}{\gamma m}.
\label{eq-mot-sx}
\end{align}
The solutions of the above equations for constant frequency and temperature are given in the appendix.

\begin{figure*}[t]
\centering
\includegraphics[width = \textwidth]{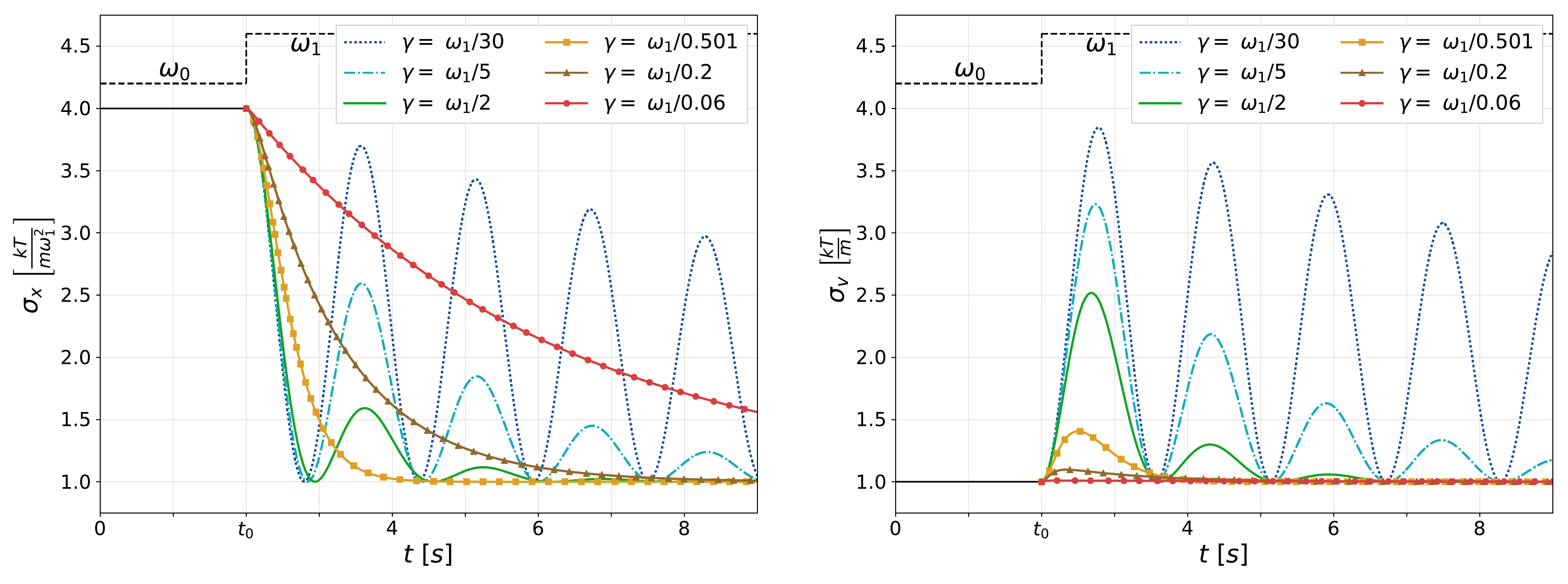}
\caption{Underdamped spatial variance $\sigma_x$  (left) and velocity variance $\sigma_v$  (right)  during the isothermal  process (9) for increasing values of the friction coefficient $\gamma$, Eqs.~\eqref{10} and \eqref{11}, respectively. Parameters are $t_0 = 2s$, $\omega_0 = 1 s^{-1}$ and $\omega_1 = 2 s^{-1}$.}
\label{plot-isothermal-var}
\end{figure*}

The stochastic heat along an individual trajectory  is respectively defined for small and large friction as \cite{sek10,sei12},
\begin{eqnarray}
Q_\text{u}&=&  \int_0^t \,dt' \langle [m\partial_{t'} v(t')+ \partial_x V(x,t')] v(t') \rangle, \label{5}\\
 Q_\text{o} &=& \int_0^t \,dt' \langle \partial_x V(x,t') v(t')\rangle\label{6}.
\end{eqnarray}
The underdamped expression \eqref{5} contains a kinetic term that accounts for the relaxation of the velocity degree of freedom. In the limit of strong friction, this thermalization is almost instantaneous and the velocity is assumed to always have its stationary value. However, this does not necessarily imply that there is no heat flow associated with the velocity relaxation \cite{hon00}. In the harmonic case, the stochastic heats \eqref{5} and \eqref{6} can be directly expressed in terms of the position and velocity variances $\sigma_x$ and $\sigma_v$,
\begin{eqnarray}
Q_\text{u} &=& \gamma kT t - \gamma m \int_0^t \, dt'\sigma_v  , \label{7}\\
Q_\text{o}&=& \frac{m}{2} \omega^2 \int_0^t \,dt' \dot{\sigma}_x  .\label{8}
\end{eqnarray}
These two equations form the basis of our study of heat leakages in stochastic  harmonic systems.

\section{Isothermal process}
We begin by investigating isothermal processes during which the frequency of the potential is varied at constant temperature. We assume that the oscillator is initially  at thermal equilibrium at $t=0$ with frequency $\omega_0$. The corresponding initial conditions for the variances are  $\sigma_{v0} = {kT}/{m}$, $\sigma_{x0} =  {kT}/{m\omega_0^2}$ and $ \dot{\sigma}_{x0} = 0$ due to the equipartition theorem. We drive the system by instantaneously changing its frequency to $\omega_1$ at time $t=t_0$,
\begin{align}
\omega(t)=\omega_0 +(\omega_1-\omega_0)\Theta(t-t_0), \label{isothermPro}
\end{align}
where $\Theta(t)$ denotes the Heavyside function.

The corresponding underdamped position and velocity variances can be obtained from Eq.~\eqref{sxv_ud}. Introducing  $\omega'=\sqrt{4\omega_1^2-\gamma^2}$  and $\tau = t-t_0$, we find,
\begin{eqnarray}
\sigma_x &=
\begin{cases}
\dfrac{kT}{m\omega_0^2}, & 0 \le t < t_0\\
\dfrac{kT}{m\omega_1^2} \bigg[1+\left(\dfrac{\omega^2_1}{\omega_0^2}-1\right) \ \dfrac{e^{-\gamma\tau}}{\omega'^2} \ (2\omega_1^2+\\
\  (2\omega_1^2-\gamma^2)\cos(\omega'\tau)+\gamma \omega' \sin(\omega'\tau))\bigg], &  t_0\le t 
\end{cases}\label{10} \\[0.3cm]
\sigma_v &=
\begin{cases}
\dfrac{kT}{m}, & \ \ \ \ \ \ 0 \le t < t_0\\
\dfrac{kT}{m} \bigg[1+\dfrac{2\omega_1^2}{\omega'^2}\left(\dfrac{\omega^2_1}{\omega_0^2}-1\right)\ e^{-\gamma\tau} \times \\[0.2cm] 
\ (1-\cos(\omega'\tau))\bigg], & \ \ \ \ \ \ t_0\le t. \label{11}
\end{cases}
\end{eqnarray}
The time dependence of the variances $\sigma_x$ and $\sigma_v$ is shown in Fig.~\ref{plot-isothermal-var} for increasing values of the friction coefficient. We observe a qualitatively different behavior in the underdamped and overdamped regimes. For small friction, the position and velocity variances settle to their respective equilibrium values,  $\sigma_{x1} =  {kT}/{m\omega_1^2}$ and $\sigma_{v1} = {kT}/{m}$, in a slow oscillatory fashion.  We note that these oscillations are out of phase, revealing the continuous conversion of kinetic to potential energy and vice versa. By contrast, for strong friction, the position variance reaches its new equilibrium value exponentially fast, while the velocity variance remains quasi constant at its initial value. We additionally emphasize that the 
relaxation time first decreases with increasing $\gamma$, before it starts increasing for higher values of  the friction coefficient.

\begin{figure}[h]
\centering
\includegraphics[width=0.48\textwidth]{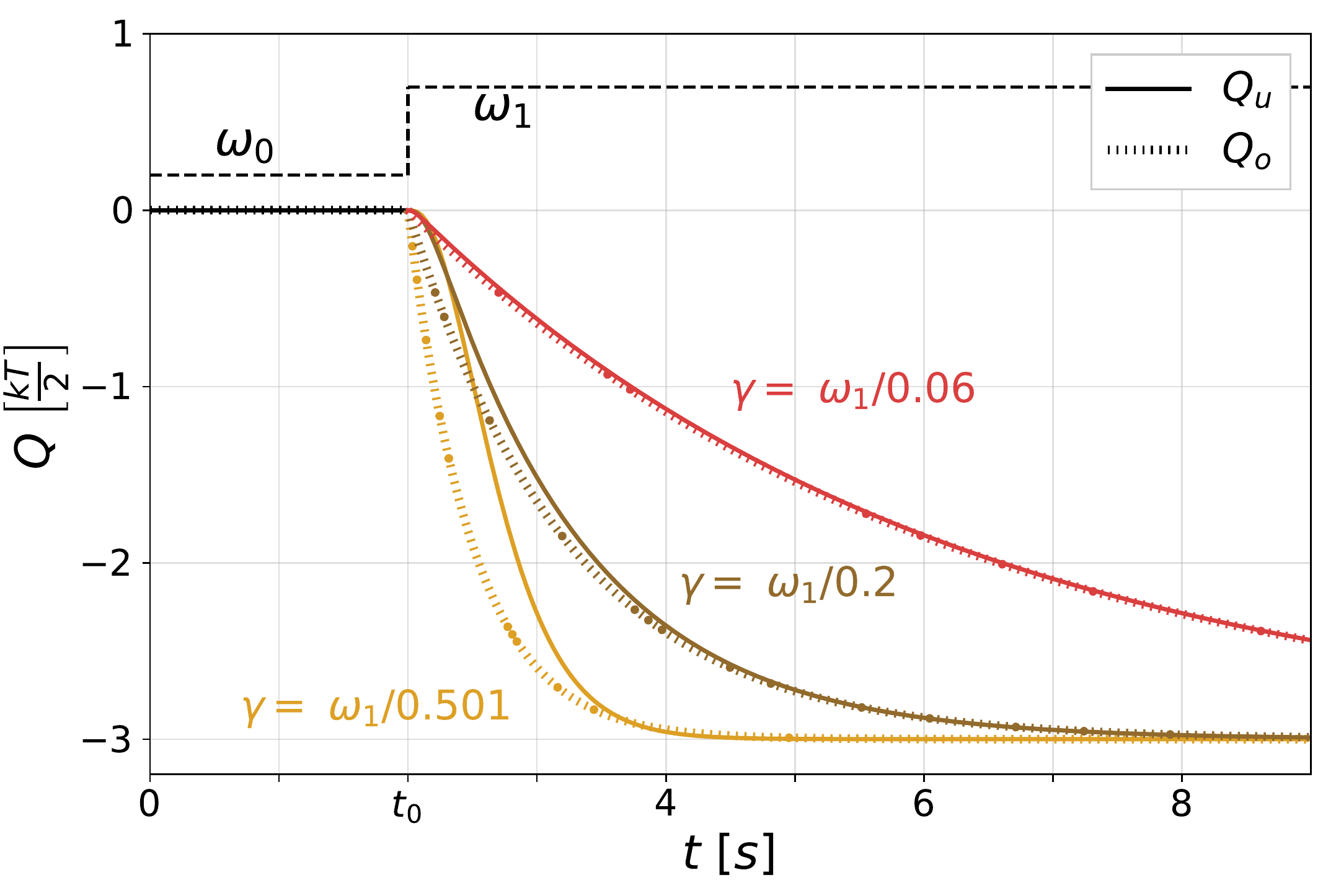}
\caption{Underdamped heat $Q_\text{u}$  (solid)  and overdamped heat $Q_\text{o}$ (dotted) during the isothermal process (9) for increasing values of the friction  coefficient $\gamma$ obtained from Eqs.~\eqref{12} and \eqref{13} respectively. Same parameters as in Fig.~1.}
\label{plot-isothermal-heat}
\end{figure}

We next compute the underdamped heat \eqref{7} using the variance \eqref{10} and the overdamped heat \eqref{8} using Eq.~\eqref{sx_od}. We obtain,
\begin{eqnarray}
Q_\text{u}& =
\begin{cases}
0, & 0 \le t < t_0\\
\dfrac{kT}{2} \left( \dfrac{\omega^2_1}{\omega_0^2}-1 \right) \bigg[\dfrac{1}{\omega'^2}e^{-\gamma \tau}(4\omega^2_1-\\
\ \gamma^2 \cos(\omega' \tau)+\gamma \omega' \sin(\omega'\tau))-1\bigg], & t_0\le t
\end{cases} \label{12}\\
Q_\text{o} &=
\begin{cases}
0,& \ \ \ 0 \le t < t_0\\
\dfrac{kT}{2}\left(\dfrac{\omega^2_1}{\omega_0^2}-1\right)\left(e^{-\tfrac{2\omega_1^2}{\gamma}\tau}-1\right), & \ \ \ t_0\le t.
\end{cases} \label{13}
\end{eqnarray}
These two formulas for the heat are displayed as a function of time for increasing values of the friction in Fig.~2. Three points are noteworthy: first, the two heats are negative, indicating that energy is given to the bath in order to compensate for the work done on the system by the sudden frequency switch; second, both expressions relax exponentially to the same value in the limit of long times, $Q_\text{u} \rightarrow  Q_\text{o}$; finally, Eqs.~(12) and (13) become identical for large friction coefficients, $\gamma \gg \omega_1$, as naively expected. The last point may be confirmed analytically by Taylor expanding $Q_\text{u}$ using $\omega' = i\gamma \sqrt{1-\alpha}\simeq i\gamma (1-\alpha/2)$ with $\alpha = {4\omega_1^2}/{\gamma^2} \ll 1$. For an isothermal process, we may thus take the large friction limit either before or after evaluating the heat.

\section{Isochoric process}
\begin{figure*}
\centering
\includegraphics[width = \textwidth]{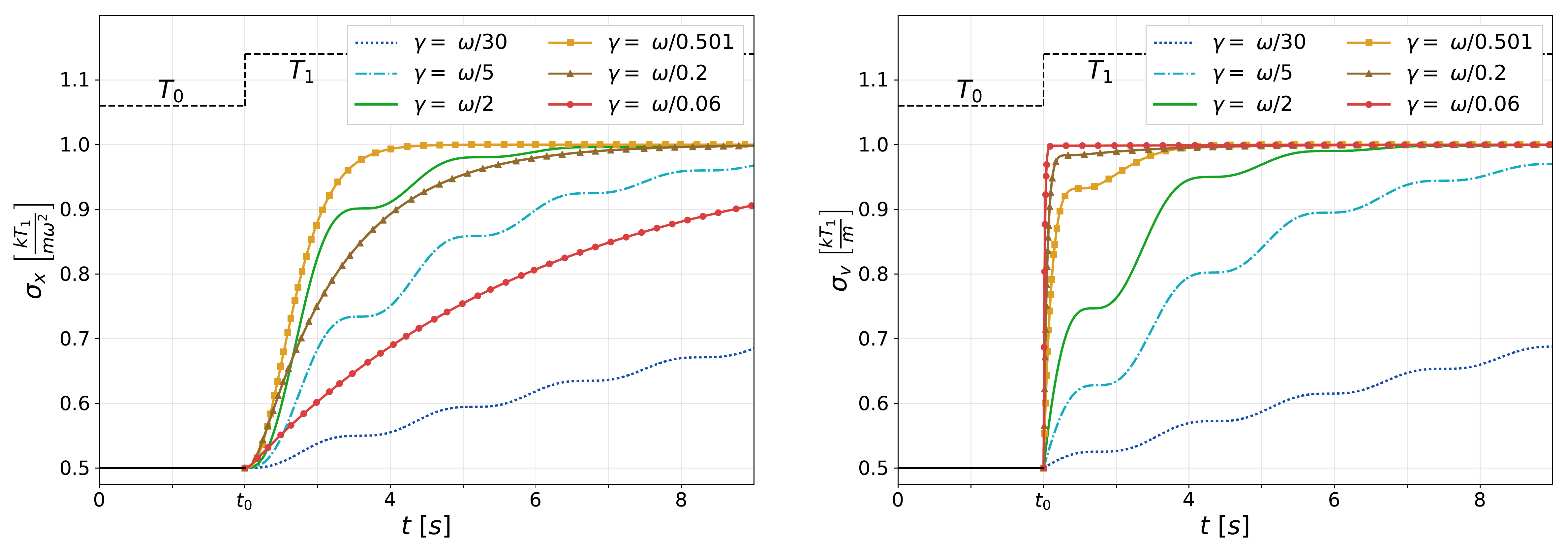}
\caption{Underdamped spatial variance $\sigma_x$  (left) and velocity variance $\sigma_v$  (right)  during the isochoric  process (14)  for increasing values of the friction coefficient $\gamma$, Eqs.~\eqref{15} and \eqref{16}, respectively. Parameters are $t_0 = 2s$, $\omega=2s^{-1}$ and $T_1/T_0 = 2$.}
\label{plot-isochoric-var}
\end{figure*}
Let us now turn to the isochoric process where the temperature is modified at constant frequency. We assume that the oscillator is initially at thermal equilibrium at temperature $T_0$ and frequency $\omega$. The corresponding initial conditions for the variances are accordingly $\sigma_{v0} = {kT_0}/{m}$, $ \sigma_{x0} =  {kT_0}/{m\omega^2}$ and $ \dot{\sigma}_{x0} = 0$. We thermally drive the system by  instantaneously switching the temperature to the value $T_1$ at $t=t_0$,
\begin{align}
T(t)=T_0 +(T_1-T_0)\Theta(t-t_0) .
\label{isochorPro}
\end{align}
The underdamped position and velocity variances  may be calculated as in the previous section.  We obtain with $\omega' = \sqrt{4\omega^2-\gamma^2}$,
\begin{eqnarray}
\sigma_x =
\begin{cases}
\dfrac{kT_0}{m\omega^2}, & \ \ 0 \le t < t_0\\[0.2cm]
\dfrac{kT_1}{m}\Big(1+\left(\dfrac{T_0}{T_1}-1\right) \dfrac{e^{-\gamma \tau}}{\omega'^{2}}[4\omega^2\\
\ -\gamma^2 \cos(\omega' \tau)+\gamma \omega' \sin(\omega'\tau)] \Big), & \ \ t_0\le t
\end{cases}\label{15}\\
\sigma_v =
\begin{cases}
\dfrac{kT_0}{m}, & \ \ 0 \le t < t_0\\[0.2cm]
\dfrac{kT_1}{m}\Big(1+\bigg(\dfrac{T_0}{T_1}-1\bigg) \dfrac{e^{-\gamma \tau}}{\omega'^{2}}[4\omega^2\\
\ -\gamma^2 \cos(\omega' \tau)-\gamma \omega' \sin(\omega'\tau)]\Big), & \ \ t_0\le t.
\end{cases}\label{16}
\end{eqnarray}
Equations (15) and (16) are shown in Fig.~\ref{plot-isochoric-var} as a function of time for increasing  friction.
The position variance $\sigma_x$ reaches its new equilibrium value, $\sigma_{x1}={kT_1}/{m\omega^2}$, exponentially with a decay time that first decreases before it increases with larger $\gamma$. On the other hand, the velocity variance $\sigma_v$ exhibits two different behaviors: for small friction, it slowly equilibrates to $\sigma_{v1}={kT_1}/{m}$, whereas it jumps almost instantaneously to that value for high friction. The assumption that the velocity variance is quasi constant in the overdamped limit is therefore verified for both isothermal and isochoric processes. However, in the latter case, it displays an initial  sudden jump induced by the temperature variation, which is neglected in the overdamped approximation. This initial jump may be physically understood  by noting that  the system adjusts instantly to the heat bath in the overdamped limit, while it adjusts immediately to changes of the external potential in the opposite underdamped limit.

\begin{figure}[h]
\centering
\includegraphics[width=0.48\textwidth]{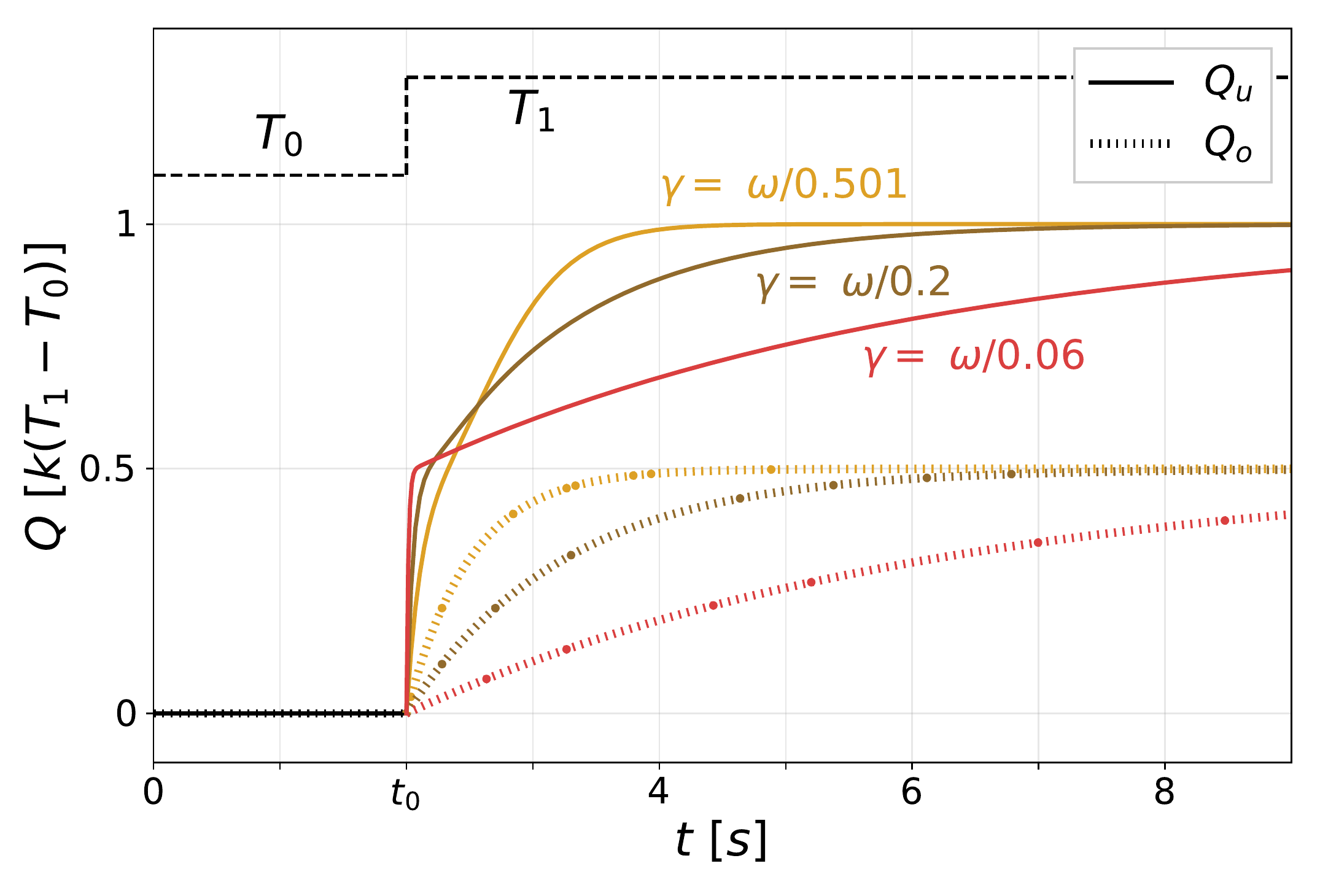}
\caption{Underdamped heat $Q_\text{u}$  (solid)  and overdamped heat $Q_\text{o}$ (dotted) during the isochoric process  (14) for increasing values of the friction coefficient $\gamma$, Eqs.~\eqref{17} and \eqref{18}, respectively. Same parameters as in Fig.~3.}
\label{plot-isochoric-heat}
\end{figure}

We may next compute the underdamped and overdamped heats in analogy to the previous section and get,
\begin{eqnarray}
Q_\text{u} &=
\begin{cases}
0, & \ \ \ 0 \le t < t_0\\
k(T_1-T_0) \bigg[ 1-\dfrac{1}{\omega'^{2}}e^{-\gamma \tau}\times \\
\ \left(4\omega^2-\gamma^2 \cos(\omega' \tau)\right)\bigg], & \ \ \ t_0\le t
\end{cases}\label{17}\\
Q_\text{o} &=
\begin{cases}
0, & 0 \le t < t_0\\
\dfrac{k(T_1-T_0)}{2}\left(1-e^{-\tfrac{2\omega^2}{\gamma}\tau}\right), & t_0\le t.
\end{cases}\label{18}
\end{eqnarray}
The two heat expressions are represented in Fig.~\ref{plot-isochoric-heat} as a function of time for increasing friction.
We first notice that the underdamped and overdamped heats differ by exactly a factor two in the long-time limit, $Q_\text{u} \rightarrow 2 Q_\text{o}$, although the work done on the system is identically zero in both situations. The equipartition theorem provides an explanation for this discrepancy. In the overdamped regime, there is only one relevant degree of freedom (the velocity being frozen). As a result, the total energies of the oscillator before and after the temperature switch are respectively ${kT_0}/{2}$ and ${kT_1}/{2}$. Since no work is done on the system, the total energy change is $\Delta E = {k(T_1-T_0)}/{2}=Q_\text{o}(t \to \infty)$. The same argument applies to the underdamped regime with now two relevant degrees of freedom (position and velocity). Consequently, $\Delta E = {k(T_1-T_0)}=Q_\text{u}(t \to \infty)$.

We additionally observe that in the limit of strong friction, the underdamped heat (17) does \textit{not} reduce to the overdamped expression (18), as was the case for the isothermal process. It further exhibits an initial jump that is directly connected to the  sudden jump of the velocity variance seen in Fig.~3. This can be seen explicitly by rewriting Eq.~\eqref{5} in the form,
\begin{equation}
 Q_\text{u} = \int_0^t dt'\, \left(\dfrac{m}{2} \dot \sigma_v +  \dfrac{m}{2} \omega^2 \dot \sigma_x \right).
\label{dQVariances}
\end{equation}
The quick relaxation of $\sigma_v$ immediately after $t_0$ thus causes a large heat flux during this short period of time, leading to the sudden jump of $Q_\text{u}$ in Fig.~\ref{plot-isochoric-heat}. The ensuing heat flux is mostly induced by the much slower relaxation of $\sigma_x$,  when $\sigma_v$ is mostly constant.
A lowest-order Taylor expansion of Eq.~(17) for $\gamma \gg \omega$ further yields,
\begin{align}
Q_\text{u} \simeq Q_\text{o} + \dfrac{k(T_1-T_0)}{2}.
\label{20}
\end{align}
The second term in Eq.~\eqref{20} is the heat leakage associated with the initial relaxation of the velocity. Its origin may  be traced  to the inertial term $m\ddot x$ in the Langevin equation (1). This term is neglected in the overdamped approximation. However, the heat leakage remains finite even for arbitrarily strong friction. This again follows from the fact that the oscillator reacts instantaneously to temperature changes  in the overdamped regime.

\section{Summary}
We have investigated the occurrence of heat leakages in harmonic systems which are  known to significantly reduce the efficiency of Brownian heat engines. Due to the conceptual simplicity of these systems, we were able to  analyze the physical origin of these heat leackages in detail and to  compute their exact expression for a sudden temperature switch  in an isochoric process. Our results emphasize the fact that the overdamped limit can be taken before or after calculating the stochastic heat for the case of a constant temperature. However, this is no longer true when the temperature changes in time, as the initial fast velocity relaxation will induce heat leakages which are not captured by the overdamped approximation. These findings complement those obtained for a spatial temperature variation in Ref.~\cite{hon00}.
 In these situations, heat should be evaluated before taking the overdamped limit.

\begin{appendix}

\section{Solutions for the variances}
\label{solvar}
We here provide for convenience the solutions of the equations (2)-(4) for the position and velocity variances for constant frequency and temperature. The underdamped equations (2) and (3) may be solved with the help of the Laplace transformation \cite{aga08}. We obtain,
\begin{equation}
\begin{aligned}
\sigma_v =& \dfrac{kT}{m} + D_1 \ e^{-\gamma t} + D_2 \ e^{(-\gamma+\omega^*)t }+D_3 \ e^{(-\gamma-\omega^*)t},\\
\sigma_x =& \dfrac{kT}{m\omega^2} + \dfrac{1}{\omega^2}e^{-\gamma t} \bigg[ D_1 + \dfrac{(\gamma + \omega^*)^2}{4\omega^2} D_2 \  e^{\omega^*t}+\\
&\dfrac{(\gamma - \omega^*)^2}{4\omega^2} D_3 \ e^{-\omega^*t} \bigg],
\label{sxv_ud}
\end{aligned}
\end{equation}
where we have defined the following quantities,
\begin{equation}
\begin{aligned}
\omega^* =& \sqrt{\gamma^2-4\omega^2} = i \omega',\\
D_1 =& \dfrac{\omega^2}{\omega^{*2}}(4\tfrac{kT}{m}-2\sigma_{v0}-2\omega^2\sigma_{x0}-\gamma \dot{\sigma}_{x0}), \\
D_2=& -\dfrac{1}{2\omega^{*2}}(\tfrac{\gamma kT}{m}(\gamma-\omega^*)+(2\omega^2-\gamma^2+\gamma \omega^*)\sigma_{v0}\\
&-2\omega^4 \sigma_{x0}+\omega^2(-\gamma + \omega^*)\dot{\sigma}_{x0}),  \\
D_3=& \dfrac{1}{2\omega^{*2}}(-\tfrac{\gamma kT}{m}(\gamma+\omega^*)+(-2\omega^2+\gamma^2+\gamma \omega^*)\sigma_{v0}\\
&+2\omega^4 \sigma_{x0}+\omega^2(\gamma + \omega^*)\dot{\sigma}_{x0}),
\label{DDD}
\end{aligned}
\end{equation}
and the initial values $\sigma_{v0}$, $\sigma_{x0}$ and $\dot{\sigma}_{x0}$.\\
On the other hand, the solution of the overdamped equation (4) is given by,
\begin{align}
\sigma_x = \frac{kT}{m\omega^2}-\left( \frac{kT}{m\omega^2} - \sigma_{x0} \right) \ e^{-2\frac{\omega^2}{\gamma}t}
\label{sx_od}
\end{align}
with the initial condition $\sigma_{x0}$.
\end{appendix}

\end{document}